\documentstyle[12pt]{article}
\begin{document}
\title{Monopoles and Solitons}
\author{B.G. Sidharth\\
Centre for Applicable Mathematics \& Computer Sciences\\
B.M. Birla Science Centre, Adarsh Nagar, Hyderabad - 500 063
(India)}
\date{}
\maketitle
\begin{abstract}
In this paper, we argue that the elusive magnetic monopole arises
due to the strong magnetic effects arising from the non
commutative space time structure at small scales.If this structure
is ignored and we work with Minkowski spacetime, then the magnetic
effect shows up as a monopole. This would also explain why the
monopole has eluded detection even after seventy years. We next
consider another area in which Solitons can be applied, viz., Bose
Einstein condensation.
\end{abstract}
\section{Introduction}
Non linear wave equations and solitons have been studied for a
long time. Let us consider some less well known application of
solitons. We start with monopoles. Ever since Dirac deduced
theoretically the existence of the monopole in 1931, it has eluded
physicists \cite{r1}. At the same time the possibility of
realising huge amounts of energy using monopoles has been an
exciting prospect. In 1980 when the fiftieth Anniversary of the
monopole was being commemorated, Dirac himself expressed his
belief that the monopole did not exist \cite{r2}. Some scholars
have indeed dismissed the monopole \cite{r3,r4}, while in a model
based on quantized vortices in the hydrodynamical formulation, the
monopole field can be mathematically identified with the momentum
vector \cite{r5}. Monopoles
had also been identified with solitons \cite{r6}.\\
In any case, it has been noted that the existence of free
monopoles would lead to an unacceptably high density of the
universe \cite{r7}, which in the light of latest observations of
an ever expanding universe
\cite{r8,r9} would be difficult to reconcile.\\
We will now show that monopoles arise due to the non commutative
structure of space time being ignored, and this would also provide
an explanation for their being undetected.
\section{The Monopole}
Let us start by reviewing Dirac's original derivation of the
Monopole (Cf.ref.\cite{r1}). He started with the wave function
\begin{equation}
\psi = Ae^{\imath \gamma},\label{e1}
\end{equation}
He then considered the case where the phase $\gamma$ in (\ref{e1})
is non integrable. In this case (\ref{e1}) can be rewritten as
\begin{equation}
\psi = \psi_1 e^{\imath S},\label{e2}
\end{equation}
where $\psi_1$ is an ordinary wave function with integrable phase,
and further, while the phase $S$ does not have a definite value at
each point, its four gradient viz.,
\begin{equation}
\kappa^\mu = \partial^\mu S\label{e3}
\end{equation}
is well defined. We use natural units, $\hbar = c = 1$. Dirac then
goes on to identify $\kappa$ in (\ref{e3}) (except for the
numerical factor $hc/e$) with the electromagnetic
field potential, as in the Weyl gauge invariant theory.\\
Next Dirac considered the case of a nodal singularity, which is
closely related to what was later called a quantized vortex (Cf.
for example ref.\cite{r10}). In this case a circuit integral of a
vector as in (\ref{e3}) gives, in addition to the electromagnetic
term, a term like $2 \pi n$, so that we have for a change in phase
for a small closed curve around this nodal singularity,
\begin{equation}
2\pi n + e \int \vec B \cdot d \vec S\label{e4}
\end{equation}
In (\ref{e4}) $\vec B$ is the magnetic flux across a surface
element $d \vec S$ and $n$ is the number of nodes within the
circuit. The expression (\ref{e4}) directly leads to the Monopole.\\
Let us now reconsider the above arguments in terms of recent
developments.\\
The Dirac equation for a spin half particle throws up a complex or
non Hermitian position coordinate. Dirac identified the imaginary
part with Zitterbewegung effects and argued that this would be
eliminated once it is realized that in Quantum Mechanics, space
time points are not meaningful and that on the contrary averages
over intervals of the order of the Compton scale have to be taken
to recover meaningful physics \cite{r11}. Over the decades the
significance of such cut off space time intervals has been
stressed by T.D. Lee and several other scholars
\cite{r12,r13,r14,r15}. Indeed with a minimum cut off length $l$,
it was shown by Snyder \cite{r16} that there would be a non
commutative space time structure, and infact at the Compton scale
we would have (Cf.ref.\cite{r15})
\begin{equation}
[x,y] = 0(l^2)\label{e5}
\end{equation}
and similar relations. The Planck scale ofcourse, is the Compton
scale for a Planck mass.\\
Infact starting from the Dirac equation itself, we can deduce
directly the non commutativity (\ref{e5}) as recently shown
\cite{r17}. This non commutative feature has also been recently
stressed, both
in Quantum Gravity and in Quantum SuperStrings theory \cite{r18,r19}.\\
Let us now return to Dirac's formulation of the monopole in the
light of the above comments. As noted above, the non integrability
of the phase $S$ in (\ref{e2}) gives rise to the electromagnetic
field, while the nodal singularity gives rise to a term  which is
an integral multiple of $2\pi$. As is well known \cite{r20} we
have
\begin{equation}
\vec \nabla S = \vec p\label{e6}
\end{equation}
where $\vec p$ is the momentum vector. When there is a nodal
singularity, as noted above the integral over a closed circuit of
$\vec p$ does not vanish. Infact in this case we have a
circulation given by
\begin{equation}
\Gamma = \oint \vec \nabla S \cdot d\vec r = \hbar \oint dS = 2\pi
n\label{e7}
\end{equation}
It is because of the nodal singularity that though the $\vec p$
field is irrotational, there is a vortex - the singularity at the
central point associated with the vortex makes the region multiply
connected, or alternatively, in this region we cannot shrink a
closed smooth curve about the point to that point. Infact if we
use the fact as seen above that the Compton wavelength is a
minimum cut off, then we get from (\ref{e7}) using (\ref{e6}), and
on taking $n = 1$,
\begin{equation}
\oint \vec \nabla S \cdot d \vec r = \int \vec p \cdot d \vec r =
2\pi mc \frac{l}{2mc} = \frac{h}{2}\label{e8}
\end{equation}
$(l = \frac{\hbar}{2mc}$ is the radius of the circuit and
$h = 2\pi$ in the above natural units).\\
In other words the nodal singularity or quantized vortex gives us
the mysterious Quantum Mechanical spin half (and other higher
spins for other values of $n$). In the case of the Quantum
Mechanical spin, there are $2 \times n/2 + 1 = n+1$ multiply
connected regions, exactly as in the case of nodal singularities.
Indeed in the case of the Dirac wave function, which is a
bi-spinor $\left(\begin{array}{ll} \Theta \\ \phi
\end{array}\right)$, it is well known that far outside the Compton
wavelength, it is the usual spinor $\Theta$, preserving parity
under reflections that predominates, whereas at and near the
Compton scale it is the spinor $\phi$ which predominates, where
under a reflection $\phi$ goes over to $-\phi$. This double
connectivity of the Dirac spinor was shown to lead immediately to
the same electromagnetic potential we had obtained from the
nonintegrability of the phase above, which again was identical to
that from Weyl's gauge invariant
theory (Cf.ref.\cite{r21} for details).\\
Let us see all this in a little greater detail \cite{r22}. We
start with a non integrable infinitessimal parallel displacement
of a four vector,
\begin{equation}
\delta a^\sigma = -\Gamma^{\sigma}_{\mu \nu} a^\mu
dx^\nu\label{e9}
\end{equation}
The $\Gamma$'s are the Christoffel symbols. This represents the
extra effect in displacements, due to curvature. In a flat space,
all the $\Gamma$'s on the right side would vanish. Considering
partial derivatives with respect to the $\mu$-th coordinate, this
would mean that, due to (\ref{e9}),
\begin{equation}
\frac{\partial a^\sigma}{\partial x^\mu} \to \frac{\partial
a^\sigma} {\partial x^\mu} - \Gamma^\sigma_{\mu \nu} a^\nu
,\label{e10}
\end{equation}
The second term on the right side of (\ref{e10}) can be written as
$$-\Gamma^{\lambda}_{\mu \nu} g^\nu_\lambda a^\sigma =
-\Gamma^{\nu}_{\mu \nu} a^\sigma ,$$ where we have linearised the
metric,
$$g_{\mu \nu} = \eta_{\mu \nu} + h_{\mu \nu},$$
$\eta_{\mu \nu}$ being the Minkowski metric and $h_{\mu \nu}$ a
small correction whose square is neglected. From (\ref{e10}) we
conclude that,
\begin{equation}
\frac{\partial}{\partial x^\mu} \to \frac{\partial}{\partial
x^\mu} - \Gamma^\nu_{\mu \nu}\label{e11}
\end{equation}
We can identify
\begin{equation}
A_\mu = \Gamma^\nu_{\mu \nu}\label{e12}
\end{equation}
from the above using minimum electromagnetic coupling exactly as
in
Dirac's monopole theory.\\
If we use (\ref{e11}), we will get the commutator relation,
\begin{equation}
\frac{\partial}{\partial x^\lambda} \frac{\partial}{\partial
x^\mu} - \frac{\partial}{\partial x^\mu} \frac{\partial}{\partial
x^\lambda} \to \frac{\partial}{\partial x^\lambda} \Gamma^\nu_{\mu
\nu}- \frac{\partial}{\partial x^\mu} \Gamma^\nu_{\lambda
\nu}\label{e13}
\end{equation}
Let us now use (\ref{e12}) in (\ref{e13}): The right side does not
vanish due to the electromagnetic field (\ref{e12}) and we have a
non-commutativity of the momentum components of quantum theory.
Indeed the left side of (\ref{e13}) can be written as
\begin{equation}
[p_\lambda , p_\mu ] \approx \frac{0(1)}{l^2},\label{e14}
\end{equation}
$l$ being the Compton wavelength. In (\ref{e14}) we have utilised
the fact that at the extreme scale of the Compton wavelength, the
Planck scale being a special case,
the momentum is $mc$.\\
From (\ref{e12}), (\ref{e13}) and (\ref{e14}), we have,
\begin{equation}
Bl^2 \sim \frac{1}{e} = \left(\frac{\hbar c}{e}\right),\label{e15}
\end{equation}
where $B$ is the magnetic field.\\
Equation (\ref{e15}) is the well-known equation for the magnetic
monopole. Indeed it has been shown by Saito and the author
\cite{r23,r22} that a non commutative spacetime at the extreme
scale shows up as
a powerful magnetic field.\\
To recapitulate, the Monopole was shown by Dirac to arise because
of two separate issues. The first was the non integrability of the
phase $S$ given in (\ref{e2}), which gave rise to the
electromagnetic potential (\ref{e3}) which was equivalent to the
Weyl potential (\ref{e12}) (which latter was dismissed because it
was adhoc). The other issue was that of nodal singularities or
alternatively the multiply connected nature of space which gave
rise to a term like $2 \pi n$ as in (\ref{e4}). In effect there
would be free monopoles. However all this was considered in the
context of the usual commutative Minkowski spacetime. Effectively
this means that terms $\sim 0(l^2)$
as in (\ref{e5}) are neglected.\\
However once such terms are included, in other words once the non
commutative structure of spacetime to this order is recognised,
firstly the previously supposedly adhoc Weyl electromagnetic
formulation automatically follows as in (\ref{e12}) and
furthermore the first term in the monopole expression (\ref{e4})
immediately gives the Quantum Mechanical spin, and the elusive
monopole appears as the magnetic effect at the Compton (or Planck
scale). Indeed in recent times the fact that non commutative
spacetime gives rise to spin has been recognized
\cite{r24,r17}.\\
\section{Non linear Equations}
Let us now come to non linear wave equation. In diverse areas of
physical application a non-linearity leads to what may be called
auto catalysis or auto production. For example in the well known
trimolecular model (Brusselator) a cubic non-linearity is required
in the rate equations\cite{r25}. In fact the non-linearity has to
be at least of the order three. The rate equations are then of the
form
$$\frac{\partial X}{\partial t} = K_1A - (K_2B+K_4)X + K_3X^2Y + D_1\nabla^2_rX,$$
and a companion equation, where the symbols have the usual significance.\\
Such a process is also well known in the upper atmosphere, in the
formation of the triple oxygen molecule, ozone\cite{r26}:
$$O + O_2 + m \to O_3 + m.$$
It is also known that the rate equations in a number of
biochemical reactions involving enzyme catalysis also exhibit in
some limiting cases, cubic terms,
for example the Glycolytic pathway (Cf.\cite{r25}).\\
In fact, in most problems of cooperative phenomena physics, for
example plasmas or lasers, at least cubic non-linearities are
required for such
cooperative behaviour.\\
In other areas, such as zoology or biology or sociology too
non-linearities cause auto production\cite{r27,r28} (Cf. for a non
technical discussion). A
very well-known example is the logistic equation.\\
There is a similar situation in quantum field theory
also\cite{r29}. It is not
surprising that auto catalysis or auto production should be a non-linear feature.\\
In fact for a linear system not only $\psi$ but also $\alpha \psi$
is a
solution.\\
However if the system is non-linear, it can be written as
$$M(\psi) = L(\psi) + N(\psi) = 0,$$
where $L$ denotes the linear part and $N$ the non-linear part. In
a first approximation, we can take
$$L(\psi_0) = 0,$$
and $N$ can be linearized by suitably substituting $\psi_0$ for
$\psi$ to get,
say, $L^{(0)}(\psi)$.\\
The system would now be approximately described by the linear
equation
$$[L + L^{(0)}] (\psi) = 0.$$
This process, if convergent can be continued. At each stage, the
coefficients of $\psi$ would depend on the linear approximation of
up to that stage. This
precisely is the characteristic of auto production.\\
A similar technique has been used recently for a Ricatti equation
derived from the Schrodinger equation of non-relativistic quantum
mechanics\cite{r30}.
\section{The Non-linear Schrodinger Equation}
In the light of the above comments, a non-linear Schrodinger
equation was deduced and used to argue that the origin of inertial
mass lies in self interacting amplitudes within, typically the
Compton wavelength\cite{r31,r32,r33,r34}. A cubic non-linearity
associated with auto catalysis is responsible for the
generation of the inertial mass.\\
In this case the equation is given by
\begin{equation}
\imath \hbar \frac{\partial \psi}{\partial t} = \frac{-\hbar^2
\partial^2 \psi} {\partial x^2} + \int H(x,x')\psi(x')
dx'\label{e16}
\end{equation}
where
\begin{equation}
H(x,x') = (\psi (x')|\psi (x)>,\label{e17}
\end{equation}
All this can be generalized immediately to the three dimensional
case.
\begin{equation}
E\psi = - \frac{\hbar^2}{2m} \nabla^2 \psi + \int
H(r,r')\psi(r')d^3r'\label{e18}
\end{equation}
 If next in an equation like (\ref{e17}), the amplitude for a
particle at $r'$ to be at $r$ vanishes outside a small interval,
so that a $\delta$ function can be introduced in (\ref{e17}), then
we have the equation
\begin{equation}
E\psi = -\frac{\hbar^2}{2m} \nabla^2 \psi + g \psi^3\label{e19}
\end{equation}
Let us now consider partial wave decomposition equation
(\ref{e18}) in spherical symmetry. This gives
\begin{equation}
[\frac{d^2}{dr^2} + k^2 -\frac{l(l+1}{r^2}], u = gu^3, k^2 =
\frac{2m}{\hbar^2} E\label{e20}
\end{equation}
Further specializing to the case $l = 0$ and $k^2 \approx 0$ and
in the spirit of the considerations in Section 1, neglecting the
cubic term in (\ref{e19}), we have in the zeroeth approximation
for $u, u \approx r$. Now writing in (\ref{e20}), $u^3$ as $u^2
\cdot u$, that is linearizing equation (\ref{e20}) and using the
zeroeth approximation we get
\begin{equation}
[\frac{d^2}{dr^2} + (k^2-r^2)] u = 0 (k^2 \approx 0)\label{e21}
\end{equation}

Equation (\ref{e21}) is the well known Harmonic Oscillator
equation with degenerate energy levels\cite{r35}. As is well
known, a set of Harmonic oscillators as above represents an
assembly of Bosons. Thus we have a collection of closely packed
nearly zero energy
Bosons similar to the Bose-Einstein condensation\cite{r36}.\\
Interestingly the link between Solitons arising from the
non-linear equations and Bose-Einstein condensation is being
investigated, for example by Khaykovich and co-workers at the ENS
Laboratory in Paris and also at the European Laboratory for
Non-linear Spectrascopy in Italy. In these experiments a
Bose-Einstein condensate of a dilute atomic gas of Lithium atoms
is used, the inter atomic interaction providing the non-linearity.
This in turn ensures the
Solitonic propagation\cite{r37}.\\
All this could have varied applications in fields ranging from
Particle Physics to Non-linear Optics, including the possibility
of high speed fibre optic communication.

\end{document}